\documentstyle[aps,12pt,manuscript,epsfig]{revtex}
\global\firstfigfalse
\begin{document}
\preprint{INJE-TP-99-2}

\title{Scattering from an AdS$_3$ bubble and an exact AdS$_3$}

\author{ H. W. Lee and Y. S. Myung }
\address{Department of Physics, Inje University, Kimhae 621-749, Korea} 

\maketitle

\begin{abstract}
We investigate the close relationship between the potential and 
absorption cross section for test fields in an AdS$_3$ 
bubble(a 5D black hole) and an exact AdS$_3$.
There are two solutions in type IIB string theory:
an AdS$_3$ bubble corresponds to the dilatonic solution, 
while an exact AdS$_3$ is the non-dilatonic solution.
In order to obtain the cross section for an AdS$_3$ bubble, we introduce 
the \{out\}-state scattering picture with 
the AdS$_3$-AFS matching procedure.
For an exact AdS$_3$, one considers the \{in\}-state scattering 
picture with the AdS$_3$-AdS$_3$ matching. Here the non-normalizable 
modes are crucially taken into account for the matching procedure.
It turns out that the cross sections for the test fields 
in an AdS$_3$ bubble take the same forms as those in an exact 
AdS$_3$. This suggests that in the dilute gas
and the low energy limits, the S-matrix for an AdS$_3$ 
bubble can be derived from an exact AdS$_3$ space.
\end{abstract}

\newpage
\section{Introduction}
There has been a great progress in string theory of the 
D1-D5 brane system with momentum modes 
along the string direction(S$^1$). This gives us a 5D black hole(M$_5$) 
with three charges $(Q_1,Q_5,Q_n)$. The first progress was achieved in the 
Bekenstein-Hawking entropy\cite{Str96PLB99}.
Apart from the success of counting the microstates of a 5D black hole
through D-brane physics, a dynamical 
consideration  becomes an  
important issue\cite{Dha96PLB51,Das97PRL417,Cal97NPB65,Kra97PRD2173}. 
This is so because the semiclassical absorption cross section (greybody factor)
for a test field arises as a 
consequence of its potential barrier surrounding the 
horizon.  That is, this is an effect of spacetime curvature.  
More precisely, it is worth noting that a semiclassical absorption cross 
section can be derived from a solution to the differential equation 
of a test field(most often, $\bar \nabla^2 \phi =0$) on the 
supergravity side. 

The AdS/CFT correspondence states that string theory in the AdS 
is dual to a conformal field theory(CFT) defined on its 
remote boundary of AdS\cite{Mal98ATMP231}.
The semiclassical limit of spacetime physics is related 
to the large N limit of the dual CFT.
A 5D black hole(M$_5\times$S$^1\times$T$^4$) 
becomes AdS$_3\times$S$^3\times$T$^4$  near horizon  but with an asymptotically
flat space(AFS)\cite{Hyu9704005}. 
Recently this is called an AdS$_3$ bubble in AFS and 
corresponds to the dilatonic solution\cite{Bal99JHEP03001}. 
In this case the matching procedure is crucial 
for obtaining an absorption coefficient
and here we need to match an AdS$_3$ bubble to AFS.

On the other hand, one obtains AdS$_3\times$S$^3\times$T$^4$ as the other 
solution to the type IIB string theory. This is an exact AdS$_3$ with
asymptotically AdS$_3$ and corresponds to the non-dilatonic 
solution\cite{Lee98PRD104013}.
Further it includes the BTZ black hole\cite{Ban93PRD1506}.
We point out that the near horizon limit of a 5D black hole yields 
the BTZ black hole, whereas this solution accomodates to the BTZ black hole 
spacetime as a whole.

It was understood that the greybody factor calculation 
makes sense when one finds an asymptotically flat 
region as in Sec.II\cite{Mal98ATMP231}.
Hence it may not be possible in an 
AdS$_3$ because there is no asymptotic state 
corresponding to particle at infinity of AdS$_3$.
However, the authors in\cite{Lee98PRD104013} calculated the greybody factor 
for a free scalar and the dilaton both in 
M$_5\times$S$^1\times$T$^4$(an AdS$_3$ bubble) and 
AdS$_3\times$S$^3\times$T$^4$(an exact AdS$_3$) within 
the type IIB supergravity.
Here in the exact AdS$_3$ calculation we choose the non-normalizable 
modes to obtain the greybody factor. 
This corresponds to the AdS$_3$-AdS$_3$ matching procedure.
This expression denotes shorthand for a certain choice of boundary 
conditions where non-normalizable modes inject flux into AdS$_3$.
It turns out that two results of a free scalar are exactly the same.
And the results for the dilaton are the same upto a factor 3.
These show a close relationship between two approaches.
More recently, vacuum correlators of the dual CFT$_4$ were expressed 
as truncated n-point functions for the non-normalizable modes in AdS$_5$. 
One can interpret this result as an S matrix of an exact AdS$_5$ arising 
from a limit of scattering from an AdS$_5$ bubble\cite{Bal99JHEP03001}.
This supports that our calculation 
based on the non-normalizable modes is correct.

In this paper we will show that the S-matrix of an AdS$_3$ 
bubble can be derived partly from an exact AdS$_3$ space.
This is one of the current issues in the AdS/CFT correspondence.
For this purpose, we investigate the close relationship 
between the potential and absorption cross section in an 
AdS$_3$ bubble and an exact AdS$_3$. 
Comparing an AdS$_3$ bubble with an exact AdS$_3$ leads to 
an assumption that the potential of an exact AdS$_3$ is the 
left-hand side of an AdS$_3$ bubble.
For this study, we introduce the \{in\} and \{out\}-state 
pictures for an AdS$_3$ bubble.
For the exact AdS$_3$ study, one needs the \{in\}-state 
as well as non-normalizable modes.
Further, we introduce the test fields for scattering analysis. 
These are in an AdS$_3$ bubble : a free scalar($\phi$) which, 
in the decoupling limit, relates 
to an (1,1) operator ${\cal O}$ in the holographically dual theory; 
two fixed scalars($\nu, \lambda$) 
to (2,2), (3,1), and (1,3) operators; two intermediate scalars($\eta, \xi$) 
to (1,2) and (2,1) operators. On the exact AdS$_3$ side, 
the test fields are a free scalar($\psi$), the dilaton($\Phi$), 
an intermediate scalar($\eta$), and the tachyon($T$).

The organization of this work is a follows. Sec.II is devoted to 
analyzing the scattering from an AdS$_3$ bubble within 
a 5D black hole. 
This corresponds to a conventional scattering study. 
We study the scattering of the test fields in an exact AdS$_3$ in Sec. III.
In this case we are careful for obtaining the absorption 
coefficient because we cannot define the asymptotically flat space.
Finally we discuss our results in Sec. IV.

\section{Scattering from an AdS$_3$ bubble in a 5D black hole}
Initially we introduce  all perturbing modes 
in a 5D black hole background.
It is pointed out that in s-wave calculation  the fixed scalars are 
physically propagating modes and other fields belong 
to be redundant modes\cite{Kra97PRD2173}.  
Hence we choose two fixed scalars $(\nu,\lambda)$ 
and a free scalar$(\phi)$ as the relevant modes.
We begin with the 5D black hole with three charges,
\begin{equation}
ds_{\rm 5D}^2= - h f^{-2/3}dt^2 +  f^{1/3}( h^{-1} dr^2 + r^2 d\Omega_3^2),
\label{metric-5D}
\end{equation}
where
\begin{equation}
f = f_1 f_2 f_n = \left (1 + { r^2_1 \over r^2}\right )
   \left (1 + { r^2_5 \over r^2}\right )
   \left (1 + { r^2_n \over r^2}\right ),~~~
h = \left (1 - { r^2_0 \over r^2}\right ).
\label{f-h}
\end{equation}
Here the radii are related to the boost parameters $(\alpha_i)$ and the charges $(Q_i)$ as
\begin{equation}
 r_i^2 = r_0^2 \sinh^2 \alpha_i = \sqrt{Q_i^2 + {r_0^4 \over 4}} - 
{r_0^2 \over 2}, i = 1, 5, n.
\label{ri}
\end{equation}
Hence the D-brane black hole depends on the 
four parameters $(r_1, r_5, r_n, r_0)$.
The event 
horizon (outer horizon) is clearly at $r=r_0$.  When all three charges are 
nonzero, the surface $r=0$ becomes a smooth inner horizon (Cauchy horizon).  
When at least one of the charges is zero, 
the surface $r=0$ becomes singular.  The 
extremal case corresponds  to the limit of $r_0 \rightarrow 0$ with the boost 
parameters $\alpha_i \rightarrow \pm \infty$, keeping the  charges 
$(Q_i)$ fixed.   We are interested in the limit of
 $r_0, r_n \ll r_1, r_5$, which is called 
the dilute gas limit. 
This is so because this limit corresponds to the near-outer horizon. 
Here we choose $Q_1 = r_1^2, Q_5 = r_5^2$, and $r_n = r_0 \sinh \alpha_n$
with a finite $\alpha_n$. This corresponds to the near-extremal black hole 
and its thermodynamic quantities 
(energy, entropy, Hawking temperature) are given by
\begin{eqnarray}
&&E_{\rm next} = {2 \pi^2 \over \kappa_5^2} \left [ r_1^2 +
r_5^2 + {1 \over 2}r_0^2\cosh 2 \alpha_n \right ],
\label{energy-next}\\
&&S_{\rm next} = { 4 \pi^3 r_0 \over \kappa_5^2} r_1 r_5\cosh \alpha_n,
\label{entropy-next}\\
&&{1 \over T_{\rm H,next}} = {2 \pi \over r_0} r_1 r_5 \cosh \alpha_n,
\label{temperature-next}
\end{eqnarray}
where $\kappa^2_5$ is the 5D gravitational constant. 
The above energy and entropy are those of a gas of massless 1D particles.  
In this case the  temperatures for left and right moving
string modes are given by
\begin{equation}
T_L = {1 \over 2 \pi} \left ( {r_0 \over r_1 r_5} \right ) e^{\alpha_n},~~
T_R = {1 \over 2 \pi} \left ( {r_0 \over r_1 r_5} \right ) e^{-\alpha_n}.
\label{T_LR}
\end{equation}
This implies that the (left and right moving) 
momentum modes along the string direction  are 
excited, while the excitations of D1-anti D1 and D5-anti D5-branes
are suppressed.  The Hawking temperature is given by 
their harmonic average
\begin{equation}
{2 \over T_H} = {1 \over T_L} + {1 \over T_R}.
\label{T_H}
\end{equation}

\subsection{Potential analysis}
For a free scalar 
$(\phi=\phi(r) e^{i\omega t} Y_l(\theta_1, \theta_2, \theta_3))$, 
the linearized equation $\bar \nabla^2 \phi=0$ 
in the background of (\ref{metric-5D}) leads 
to\cite{Dha96PLB51,Das97PRL417,Mal97PRD4975}
\begin{equation}
\left [ \left ( h r^3 \partial_r\right )^2 + \omega^2 r^6 f - 
  {  l(l+2) h\over r^2  }
   \right ] \phi = 0.
\label{eq-free}
\end{equation}
The s-wave($l=0$) linearized equation
for the fixed scalars takes the form\cite{Cal97NPB65,Kra97PRD2173}
\begin{equation}
\left [ \left ( h r^3 \partial_r\right )^2 + \omega^2 r^6 f - 
  {{8 h r^4 r_{\pm}^4} \over (r^2 + r_{\pm}^2)^2 }
  \left ( 1 + {r_0^2 \over r_{\pm}^2} \right ) \right ] \phi_{\pm}=0,
\label{eq-swave}
\end{equation}
where one gets $\nu$ for $\phi_+$ and $\lambda$ for $\phi_-$.
Here $r_\pm^2= [ r_1^2 + r_5^2 + r_n^2 
\pm \sqrt{r_1^4 + r_5^4 + r_n^4 -r_1^2 r_5^2 -r_1^2 r_n^2 
-r_5^2 r_n^2 } ]/3$.
Considering $ N = r^{-3/2} \tilde N$, 
for $N = \nu, \lambda, \phi$ and 
introducing a tortoise coordinate 
$ r^* = \int {(dr/h)} = r + (r_0/2)\ln |(r - r_0)
/(r + r_0)|$\cite{Dha96PLB51}, then the equation takes the form
\begin{equation}
{d^2 \tilde N \over d r^{*2} } 
+ (\omega^2 - \tilde V_N) \tilde N = 0. 
\label{eqom}
\end{equation}
Here we take $r_1 = r_5 = R$ and $r_0 = r_n$ for simplicity.
In the dilute gas limit $(R \gg r_0), \tilde V_N(r)$ is given by
\begin{equation}
\tilde V_\nu(r) = - \omega^2 (f -1)  
+ h \left [ { 3   \over 4 r^2 } \left (1 + { 3 r_0^2 \over r^2} \right ) + 
{ 8 R^4  \over r^2 (r^2 + R^2)^2 } \right ],
\label{potential-nu}
\end{equation}
\begin{equation}
\tilde V_\lambda(r) = - \omega^2 (f -1)   
+ h \left [ { 3   \over 4 r^2 } \left (1 + { 3 r_0^2 \over r^2} \right ) + 
{ 8 R^4  \over r^2 (3 r^2 + R^2)^2 }\right ],
\label{potential-lambda}
\end{equation}
\begin{equation}
\tilde V_\phi(r) = - \omega^2 (f -1)  
+ h \left [ { 3  \over 4 r^2 } \left (1 + { 3 r_0^2 \over r^2} \right ) + 
{ l(l+2) \over r^2  }\right ],
\label{potential-phi}
\end{equation}
where 
\begin{equation}
f -1 = { r_0^2 + 2 R^2 \over r^2} + { (2 r^2_0 + R^2) R^2 \over r^4}
        + { r^2_0 R^4 \over r^6}.
\end{equation}

\begin{figure}
\epsfig{file=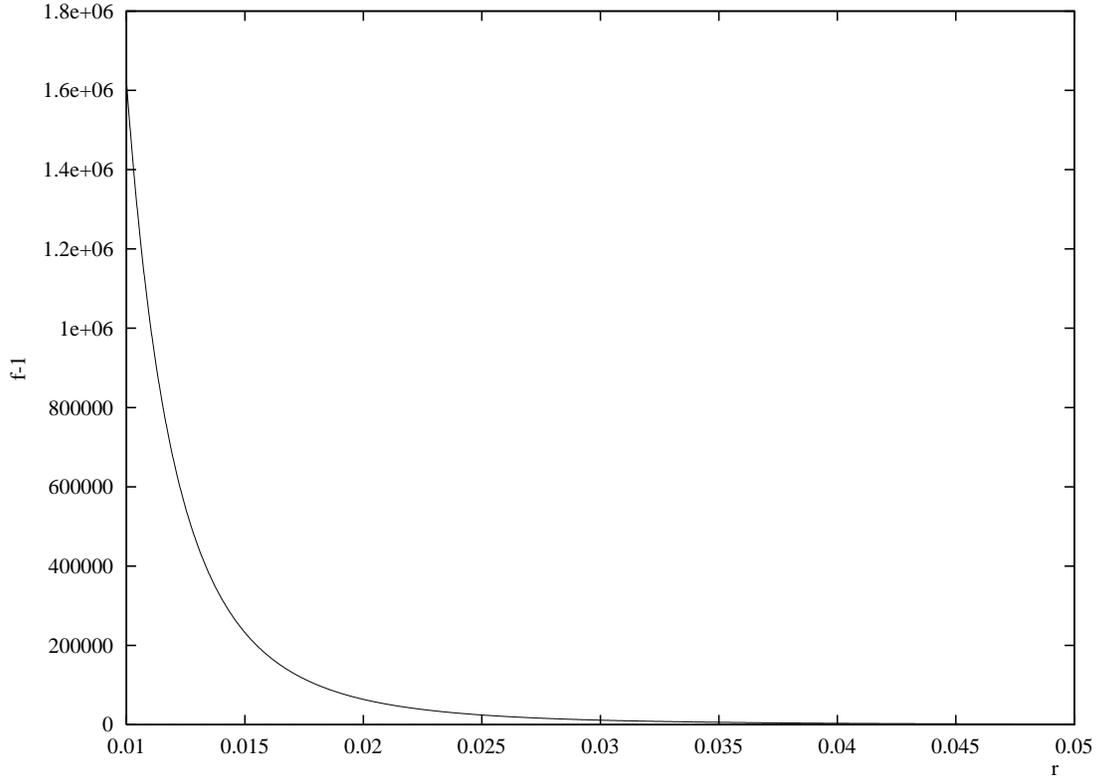,width=0.9\textwidth,clip=}
\caption{
\label{fig-f-1}
The graph of $(f-1)$ in 5D black hole with $r_0=0.01, R=0.3$. 
A peak appears at outer horizon($r=r_0$).}
\end{figure}

We note that $\tilde V_N$ depends on two parameters ($r_0, R$)
as well as the energy$(\omega)$. As (\ref{eqom}) stands, it is far from
the Schr\"odinger-type equation. The $\omega$-dependence is 
a matter of peculiar interest
to us compared with the Schwarzschild black hole 
potentials $(V_{RW}, V_Z, V_\psi)$\cite{Reg57PR1403}. This makes the 
interpretation of $\tilde V_N$ as a potential difficult. 
As is shown in Fig.\ref{fig-f-1}, this arises 
because $(f-1)$ is very large as $10^6$ for $r_0 = 0.01,
R= 0.3 $ in the near-horizon.
In order for $\tilde V_N$ to be a potential, 
it is necessary to take the low energy limit of $\omega \to 0$.
It is suitable to be $10^{-3}$.  
And $\omega^2(f-1)$ is of order ${\cal O}(1)$ and thus
it can be ignored in comparison to the remaing ones. 
Now we can define a potential 
$V_N = \tilde V_N + \omega^2 ( f- 1)$.
Hence, in the low energy limit($\omega \to 0$), Eq.(\ref{eqom}) 
becomes as the Schr\"odinger-type equation.
Further the last terms in (\ref{potential-nu})-(\ref{potential-lambda}) 
are important to compare each other.
After the partial fraction, these lead to
\begin{eqnarray}
&& { 8 R^4   \over r^2 (r^2 + R^2)^2 }= { 8 \over r^2} - { 8 \over r^2 + R^2}
- { 8 R^2 \over (r^2 + R^2)^2},\label{par-far1}\\
&&{ 8 R^4   \over r^2 (3 r^2 + R^2)^2 }= { 8 \over r^2} - 
{ 24 \over 3 r^2 + R^2}
- { 24 R^2 \over (3 r^2 + R^2)^2}.
\label{par-far2}
\end{eqnarray}
The last term of 
a free scalar in (\ref{potential-phi}) with $l=2$ keeps the first
terms  in (\ref{par-far1})-(\ref{par-far2}) only. 
One finds immediately the sequence 
\begin{equation}
 V_{\phi_0} \ll V_\lambda \leq V_\nu \leq V_{\phi_2}.
\label{potential-seq}
\end{equation}
\begin{figure}
\epsfig{file=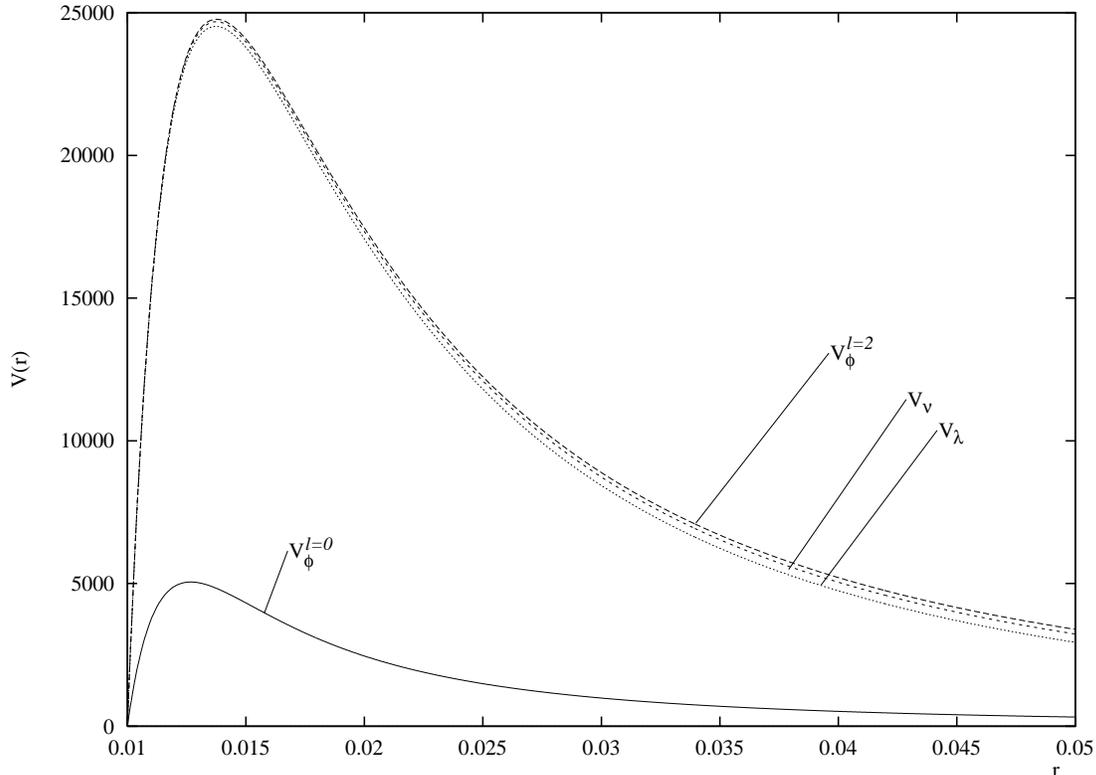,width=0.9\textwidth,clip=}
\caption{
\label{fig-potential}
Four potential graphs ($V_\phi^{l=0}, V_\nu, V_\lambda, V_\phi^{l=2}$)
for an AdS$_3$ bubble in a 5D black hole with
$r_0=0.01, R=0.3$.
}
\end{figure}
\noindent
Here $\phi_0$ denotes the s-wave($l=0$) free scalar and
$\phi_2$ the  free one with $l=2$.
This is also observed from the graphs of potential in Fig.\ref{fig-potential}
with $ r_0 = 0.01, R = 0.3$.
Because the shape of their potentials takes nearly the same form 
($V_\lambda \simeq V_\nu \simeq V_\phi^{l=2}$), these 
give us nearly the same reflection coefficient ${\cal R}=|R|^2$ 
and absorption one ${\cal A}=|A|^2$.  
For example,  in the low energy limit of 
$\omega \to 0$, $\lambda, \nu, \phi_2$ take 
nearly the zero-absorption cross section. 
Furthermore all potentials go to zero, as $r$ approaches infinity.
This implies the existence of the asymptotic states outside an AdS$_3$ bubble.
The size of an AdS$_3$ bubble is from $r_0=0.01$ to $R=0.3$.

\subsection{Scattering from an AdS$_3$ bubble}
\begin{figure}
\epsfig{file=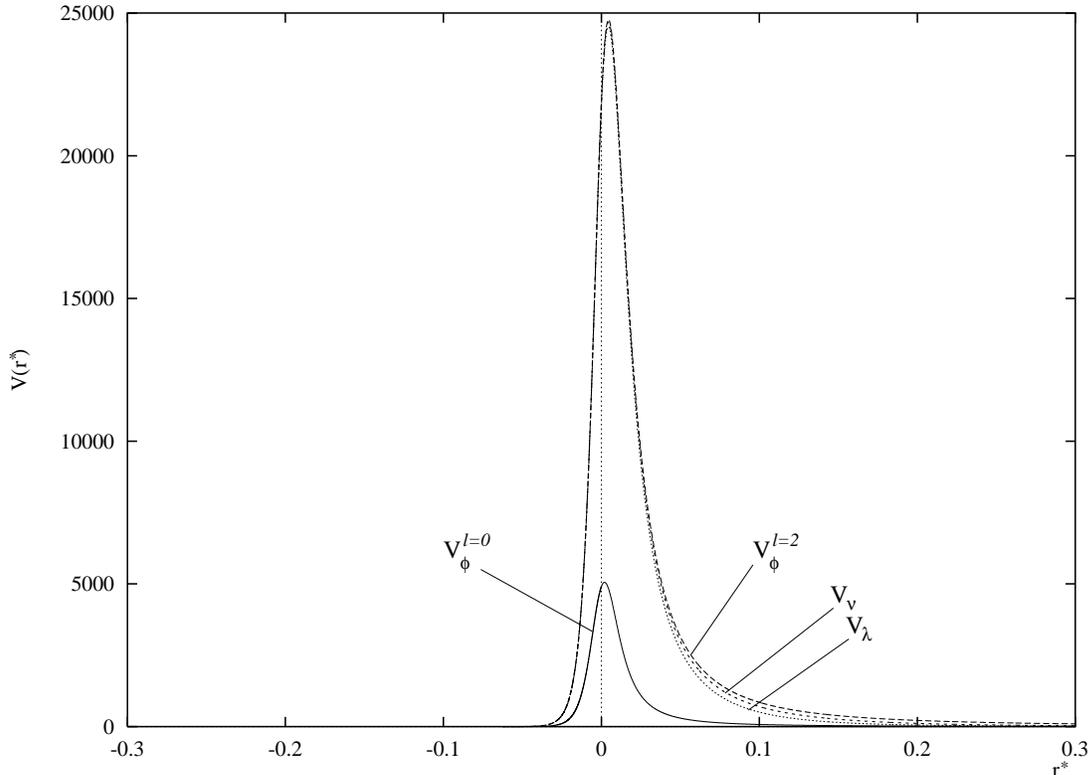,width=0.9\textwidth,clip=}
\caption{
\label{fig-potential-AFS-tor}
Four potential graphs ($V_\phi^{l=0}, V_\nu, V_\lambda, V_\phi^{l=2}$) 
as functions of $r^*$
for an AdS$_3$ bubble in a 5D black hole with $r_0=0.01, R=0.3$.
}
\end{figure}

We are interested in the scattering of the test fields off 
$V_N(r^*)$. It is well known that the scattering analysis 
is usually done by choosing a coordinate such as $-\infty \le r^* \le \infty$.
It is always possible to visualize the black hole as presenting 
an effective potential barrier (or well) to an incoming test wave. 
One expects that some of the incident wave will be irreversibly 
absorbed by the black hole, while the remaining fraction will be 
scattered back to the infinity. 
In this scattering we can calculate the reflection and 
transmission(absorption) coefficients\cite{Dij92NPB269}. 
As is shown in Fig.\ref{fig-potential-AFS-tor}, 
we note that all potential barriers $V_N(r^*)$ take nearly the symmetric 
forms around $r^*=0$ when they are rewritten by a tortoise 
coordinate $r^*$\cite{Cha83}.
Also they are localized at $r^*=0$.
Since $V_N(r^*) \to 0$ as $r^* \to \pm \infty$, one finds 
\begin{equation}
{ d^2 \tilde N_{\pm\infty} \over {d r^*}^2 } +
\omega^2 \tilde N_{\pm\infty} =0.
\label{eq-Ntilde}
\end{equation}
Asymptotically($r^*\to \infty$) the solution is given by
\begin{equation}
N_{+\infty}^{\rm out} 
 = e^{i \omega r^*} + R_N^{\rm out}(\omega) e^{-i \omega r^*}.
\label{sol-N-asymp}
\end{equation}
Considering the time-part of $e^{i \omega t}$, the first 
is an incoming wave($\leftarrow$) and the last is 
an outgoing wave($\to$).
Near the horizon it is purely incoming($\leftarrow$) as 
\begin{equation}
N_{-\infty}^{\rm out} = A_N^{\rm out}(\omega) e^{i \omega r^*}.
\label{sol-N-near}
\end{equation}
We call this type of solution as $\{{\rm out} \}_N$ and the 
corresponding vacuum state is defined as $b_i | 0 \rangle_{\rm out} =0$. 

In order to study the Hawking radiation,  we introduce another 
boundary condition.
Asymptotically the wave is purely outgoing($\to$)
\begin{equation}
N_{+\infty}^{\rm in} = A_N^{\rm in}(\omega) e^{-i \omega r^*}
\label{sol-Nin-asymp}
\end{equation}
but near the horizon it has both outgoing($\to$) and 
incoming($\leftarrow$) parts as
\begin{equation}
N_{-\infty}^{\rm in} = e^{-i \omega r^*} + 
       R_N^{\rm in}(\omega) e^{i \omega r^*}.
\label{sol-Nin-near}
\end{equation}
We call this type of solution as $\{ {\rm in} \}_N$ and its 
vacuum state is defined as $a_i | 0 \rangle_{\rm in} =0$.
The vacuum states $|0\rangle_{\rm out}$ and $|0\rangle_{\rm in}$ form 
two different bases of which any state can be 
expanded in terms of the other. These are 
two different Fock space vacuum states and $\{{\rm out}\}_N$ and 
$\{ {\rm in} \}_N$ are related to each other by the 
Bogoliubov transformation,
\begin{eqnarray}
b_i &=& \sum_j ( \alpha_{ij}^* a_j - \beta_{ij}^* a_j^\dagger ),
\label{Bog-b} \\
b_i^\dagger &=& \sum_j ( \alpha_{ij} a_j^\dagger - \beta_{ij} a_j ).
\label{Bog-bdagger} 
\end{eqnarray}
The computation of Hawking shows in a semi-classical 
approximation that
the thermal radiation from the black hole with temperature $T_H$ 
is given by\cite{Hwk75CMP199}
\begin{equation}
\langle N_\omega \rangle =
\raisebox{-0.7ex}{{\rm\scriptsize in}}
\langle 0 | b_i^\dagger b_i | 0 \rangle
\raisebox{-0.7ex}{{\rm\scriptsize in}}
= \sum_k | \beta_{ik} |^2 =
{ \sigma_{\rm 5D}^N \over {e^{\omega/T_H} -1 }} ,
\label{thr-rad}
\end{equation}
with an absorption cross section 
$\sigma_{\rm 5D}^N = |A_N^{\rm out} |^2 \times 4 \pi / \omega^3 =
{\cal A}_N^{\rm out} 4 \pi / \omega^3$.
Note that if $\sigma_{\rm 5D}^N$ is a constant, 
$\langle N_\omega \rangle$ is the same as that of a black body.
Typically, $\sigma_{\rm 5D}^N$ is not constant but varies.
The deviations from the black body spectrum have earned it the 
name ``greybody factor''.
Here we define the S-matrix from $\{{\rm in}\}_N$ and $\{ {\rm out}\}_N$ 
as\cite{Cha83}
\begin{equation}
S(\omega) =
\left (
\begin{array}{cc}
A_N^{\rm out}(\omega) & A_N^{\rm in}(\omega)
 \\
R_N^{\rm out}(\omega) & R_N^{\rm in}(\omega)
\end{array}
\right ).
\label{S-matrix}
\end{equation}
Here an incident wave($e^{i \omega r^*}$) of unit amplitude from 
$r^*= + \infty$ gives rise to $A_N^{\rm out}(\omega)$ 
and $R_N^{\rm out}(\omega)$.
On the other hand, an incident wave($e^{- i \omega r^*}$) of 
unit amplitude from $r^* = -\infty$ gives rise to $A_N^{\rm in}(\omega)$
and $R_N^{\rm in}(\omega)$.
The relation between these is given by
\begin{equation}
A_N^{\rm out}(\omega) = A_N^{\rm in}(\omega) = A_N(\omega),
\label{relation-A}
\end{equation}
\begin{equation}
{R_N^{\rm out}(\omega) \over A_N(\omega)} =
-{R_N^{\rm in}(-\omega) \over A_N(-\omega)} , ~~
{R_N^{\rm out}(-\omega) \over A_N(-\omega)} =
-{R_N^{\rm in}(\omega) \over A_N(\omega)} , 
\label{relation-R}
\end{equation}
\begin{equation}
A_N^*(\omega) = A_N(-\omega),~~
{R_N^{\rm out}}^*(\omega) = R_N^{\rm out}(-\omega),~~
{R_N^{\rm in}}^*(-\omega) = R_N^{\rm in}(\omega).
\label{relation-reverse}
\end{equation}
The above relations establish the symmetry and unitarity of the 
S-matrix in the AdS$_3$ bubble scattering.
Actually  $A_N(\omega)$ can be calculated from the backscattering 
of an incident wave $N$ off the potential $V_N(r^*)$.
It is not easy to find out the absorption amplitude
$A_N$ directly in the complicated potentials such as $V_N(r^*)$. 
In this case one use the flux ${\cal F}$ of the incoming wave 
to obtain the absorption coefficient
\begin{equation}
{\cal A}_N^{\rm out} = { {\cal F}(-\infty) \over {\cal F}(\infty)},
\label{abs-coef}
\end{equation}
where ${\cal F}(-\infty) ({\cal F}(\infty))$ are 
the fluxes at the horizon(infinity).
In this way we can calculate the semiclassical 
absoprtion cross section.

\subsection{Exact analysis of near-horizon (AdS$_3$ bubble) }
First let us consider a free scalar. Using $z=h$, the wave 
equation (\ref{eq-free}) can be rewritten as
\begin{equation}
z(1-z) {d^2 \phi \over dz^2} + (1-z) {d \phi \over dz } +
\left \{ -C + {Q \over z} + {E \over 1-z} \right \} \phi =0,
\label{near-eq}
\end{equation}
where
\begin{eqnarray}
C&=& \left ( { \omega r_1 r_5 r_n \over 2 r_0^2} \right )^2
= { \omega^2 \over 64 \pi^2} \left ( { 1 \over T_L} - { 1 \over T_R} 
\right )^2,
 \label{coef-C} \\
E&=& - {l(l+2) \over 4} + {\omega^2 (r_1^2+r_5^2 +r_n^2) \over 4}
   \simeq - {{l(l+2)} \over 4} 
\label{coef-E} \\
Q&=& \left( {\omega \over 4 \pi T_H} \right )^2 
 \left [ \left ( 1 + {r_0^2 \over r_1^2} +{r_0^2 \over r_5^2} \right ) 
 + 4 \pi^2 r_n^2 T_H^2 \right ]
 \simeq \left ( { \omega \over 4 \pi T_H} \right )^2
 \label{coef-Q}.
\end{eqnarray}
Here $\simeq$ means both the dilute gas limit($r_0, r_n \ll r_1, r_5$) and 
low energy limit($\omega \to 0$). Then one has 
$\omega r_0, \omega r_n \ll \omega r_1, \omega r_5 < 1$.
In order to compare (\ref{near-eq}) with the hypergeometric 
equation, one has to transform it into the pole-free equation.
With an unknown constant $A$, we find the ingoing mode at the horizon
\begin{equation}
\phi^n = A z^{-i \sqrt{Q}} (1-z)^{(1-\nu)/2} F(a,b,c;z),
\label{near-sol}
\end{equation}
where
\begin{eqnarray}
\nu &=& \sqrt{(l+1)^2 - \omega^2 (r_1^2 + r_5^2 +r_n^2)} \simeq l+1,
 \label{hyper-nu} \\
a &=& {1 -\nu \over 2} - i \sqrt{Q} + i \sqrt{C}
 \simeq  - {l \over 2} - i { \omega \over 4 \pi T_R},
 \label{hyper-a} \\
b &=& {1 -\nu \over 2} - i \sqrt{Q} - i \sqrt{C}
 \simeq  - {l \over 2} - i { \omega \over 4 \pi T_L},
 \label{hyper-b} \\
c &=& 1 - 2 i \sqrt{Q} \simeq 1  - i { \omega \over 2 \pi T_H}.
 \label{hyper-c}
\end{eqnarray}
The large $r$-behavior ($z\to 1$) of $\phi^n$ can be obtained 
from the ($z \to 1-z$) transformation
rule for the hypergeometric functions as
\begin{eqnarray}
\phi^{n \to f} &=&
{{A \Gamma(c) \Gamma(c-a-b)} \over
   {\Gamma(c-a) \Gamma(c-b)}} u^{\nu -1}
+
{{A \Gamma(c) \Gamma(a+b-c)} \over
   {\Gamma(a) \Gamma(b)}} u^{-(\nu +1)}.
\label{near-far-sol}
\end{eqnarray}
For the fixed scalars($\nu, \lambda$), considering both (\ref{par-far1}), 
(\ref{par-far2}) and the near-horizon condition of 
$r \simeq r_0 \ll r_1, r_5$, one finds  that (\ref{eq-swave}) leads to 
(\ref{eq-free}) with $l=2$. Thus one can obtain their near-horizon 
behaviors from $\phi_2$.

\subsection{Asymptotic states}
Let us first consider a free scalar.
In the far region, we introduce $\phi = \breve \phi/r$ 
and $u = \omega r$ and then (\ref{eq-free}) leads to
\begin{equation}
{{d^2 \breve \phi} \over d u^2} +
{1 \over u} {d \breve \phi \over d u} +
\left [ 1 - {\nu^2 \over u^2} \right ] \breve \phi =0.
\label{far-eq}
\end{equation}
The solution is given by the Bessel function when $\nu$ is
not an integer
\begin{equation}
\phi^f = \left [ \alpha { J_\nu(u) \over u} +
   \beta {J_{-\nu} (u) \over u} \right ],
\label{far-sol}
\end{equation}
where $\alpha, \beta$ are unknown constants.
From the large $u$-behavior($r \to \infty, \omega r \gg 1$), 
one finds the asymptotic states
\begin{equation}
\phi_\infty^f = \sqrt{ 1 \over 2 \pi} 
 { e^{-i u} \over u^{3/2} } \left \{ 
 \alpha e^{i(\nu +1/2)\pi/2} + \beta e^{-i(-\nu +1/2)\pi/2}
\right \}
\label{asymp-state}
\end{equation}
and its incoming flux
\begin{eqnarray}
{\cal F}(\infty)
&=& - { 2 \over \omega^2} \left \vert
\alpha e^{i(\nu + 1/2) \pi/2} +
\beta e^{i(-\nu + 1/2) \pi/2}
\right \vert^2.
\label{far-flux}
\end{eqnarray}
The small $u$-behavior($\omega r < 1$) of $\phi^f$ is
\begin{equation}
\phi^{{\rm f} \to {\rm inter}} = { 1 \over u} \left [
\alpha \left ( { u \over 2} \right )^\nu { 1 \over \Gamma(\nu +1) } +
\beta \left ( { u \over 2} \right )^{-\nu} { 1 \over \Gamma(-\nu +1) }
\right ].
\label{far-small-sol}
\end{equation}
On the other hand, the asymptotic behavior of the s-wave fixed scalars
($ \phi_\pm = \breve \phi_\pm/r$) is governed by
\begin{equation}
{{d^2 \breve \phi_\pm} \over d u^2} +
{1 \over u} {d  \breve \phi_\pm \over d u} +
\left [ 1 - {\nu^2 \over u^2} \right ]  \breve \phi_\pm =0
\label{far-eq-swave}
\end{equation}
with $\nu \simeq 1$.  Its solution is given by
\begin{equation}
\phi_\pm^f \simeq \left [ \alpha_\pm {J_\nu(u) \over u} +
                   \beta_\pm {J_{-\nu}(u) \over u} \right ].
\label{far-sol-swave}
\end{equation}
Here $\alpha_\pm, \beta_\pm$ are unknown constants.

\subsection{AdS$_3$-AFS matching procedure and absorption cross section}
Here we use the matching procedure between 
an AdS$_3$ bubble (near-horizon of a 5D black hole) and asymptotically 
flat space to obtain an absorption coefficient.
First consider the matching of a free scalar.
Here the matching point resides on $0 < r^* < \infty$ 
in Fig.\ref{fig-potential-AFS-tor}.
In the intermediate zone ($u < 1$), from (\ref{near-far-sol}) and 
(\ref{far-small-sol}) one finds
\begin{equation}
\alpha = A u_0
{{ 2^\nu \Gamma(\nu+1) \Gamma(\nu)\Gamma(c) u_0^{-\nu}} \over
   {\Gamma(c-a) \Gamma(c-b)}},
\beta = A u_0
{{ \Gamma(-\nu+1)\Gamma(-\nu)\Gamma(c) u_0^{\nu}} \over
   {2^\nu \Gamma(a) \Gamma(b)}}.
 \label{alpha-beta-value}
\end{equation}
Since $u_0=\omega r_0 \ll 1$, one finds $ \alpha \gg \beta$.
In this case we take an incoming flux effectively. 
Furthermore the incoming flux at the horizon is found as
\begin{equation}
{\cal F}(0) = - 8 \pi r_0^2 \sqrt{Q} | A |^2.
\label{near-flux}
\end{equation}
The absoprtion coefficient is given by
\begin{equation}
{\cal A}_\phi^{\rm out} = {{\cal F}(0) \over {\cal F}(\infty)} \simeq
4 \pi u_0^2 \sqrt{Q} \left \vert {A \over \alpha} \right \vert ^2 .
\label{abs-coeff}
\end{equation}
The absorption cross section takes the form\cite{Mal97PRD4975,Mat98NPB204}
\begin{eqnarray}
\sigma^{\phi}_{\rm 5D} &=& (l+1)^2 { 4 \pi \over \omega^3}
{\cal A}_\phi^{\rm out}
\nonumber \\
&\simeq&
{ {\cal A}_H^{\rm 5D} \over [l! (l+1)!]^2 }
(l+1)^2 \left({\omega r_0 \over 2} \right )^{2l}
\left \vert
   {{\Gamma({l+2\over 2} - i {\omega \over 4 \pi T_L})
     \Gamma({l+2\over 2} - i {\omega \over 4 \pi T_R})} 
     \over
    {\Gamma(1 - i { \omega \over 2 \pi T_H})}}
\right \vert ^2
\label{abs-cross}
\end{eqnarray}
with the area of horizon ${\cal A}_H^{\rm 5D} = 2 \pi r_1 r_5 r_n$.
We have, for even $l$
\begin{eqnarray}
\sigma^{\phi_l}_{\rm 5D} 
&=& (l+1)^2 {  \pi^3 \over 2^{4l}}
{ {(r_1 r_5)^{2l +2} \omega^{2l +1} } \over 
    [ l! (l+1)! ]^2 } 
[\omega^2 + (2 \pi T_L)^2 2^2] \cdots [\omega^2 + (2 \pi T_L)^2 l^2] \times
\nonumber \\
&&
~~~~~~[\omega^2 + (2 \pi T_R)^2 2^2] \cdots [\omega^2 + (2 \pi T_R)^2 l^2] 
{{ e^{\omega/T_H} -1} \over 
 {(e^{\omega / 2 T_L} -1 ) ( e^{\omega / 2 T_L} -1 ) } }.
\label{abs-cross-l}
\end{eqnarray}
Matching procedure for the s-wave $\phi_\pm$ 
is nearly the same as in a free scalar\cite{Kra97PRD2173}. It leads to
\begin{eqnarray}
\sigma^{\phi_\pm}_{\rm 5D} 
&=& {  \pi^3 r_1^6 r_5^6 \over 64 r_\pm^4} \omega 
(\omega^2 + 16 \pi^2 T_L^2 ) (\omega^2 + 16 \pi^2 T_R^2 )
{{ e^{\omega/T_H} -1} \over 
 {(e^{\omega / 2 T_L} -1 ) ( e^{\omega / 2 T_L} -1 ) } }.
\label{abs-cross-swave}
\end{eqnarray}
In the limit of $\omega \ll T_L, T_R, T_H$, the low-energy absorption 
cross sections are
calculated as
\begin{eqnarray}
&& \sigma_{\rm 5D}^{\phi_0}  = {\cal A}_H^{\rm 5D},
\label{sigma-phi0} \\
&&\sigma_{\rm 5D}^{\phi_2}  = 
{3 \over 16} {\cal A}_H^{\rm 5D} (\omega r_0)^4 = {3 \over 4} (\omega R)^4
   \left \{{{\cal A}_H^{\rm 5D} \over 4}\left ({r_0 \over R}\right )^4 
   \right \} ,
\label{sigma-phi2} \\
&& \sigma_{\rm 5D}^{\nu}  = 
{{\cal A}_H^{\rm 5D} \over 4}\left ({r_0 \over R}\right )^4,
\label{sigma-nu} \\
&&\sigma_{\rm 5D}^{\lambda}  = 
9{ {\cal A}_H^{\rm 5D} \over 4}\left ({r_0 \over R}\right )^4,
\label{sigma-lambda}
\end{eqnarray}
where we impose the relation $r_1 = r_5 = R, r_0 = r_n$.
Here we find a sequence of cross section
\begin{equation}
\sigma_{\rm 5D}^{\phi_0} \gg \sigma_{\rm 5D}^{\lambda} \geq 
\sigma_{\rm 5D}^{\nu} \geq \sigma_{\rm 5D}^{\phi_2}.
\label{sigma-seq}
\end{equation} 
This originates from the potential sequence in (\ref{potential-seq}).
It is consistent with our naive expectation that
the absorption cross section increases, as the height of potential decreases.
Here we wish to point out the difference 
between a free scalar and the fixed scalars.
In the dilute gas limit ($ R \gg r_0$) and the low energy  
limit ($\omega \to 0$),
the s-wave cross section for a free scalar($\sigma_{5D}^{\phi_0}$) 
goes to ${\cal A}_H^{\rm 5D}$\cite{Das97PRL417},
while the s-wave cross sections for fixed scalars ($\nu,\lambda$) 
including $\phi_2$ approach zero\cite{Cal97NPB65}. 
Also this can be confirmed from Fig.\ref{fig-potential}.

\section{Scattering from an exact AdS$_3$ in 
AdS$_3$(BTZ black hole) $\times$ S$^3 \times $T$^4$}
\subsection{Potential analysis}
\begin{figure}
\epsfig{file=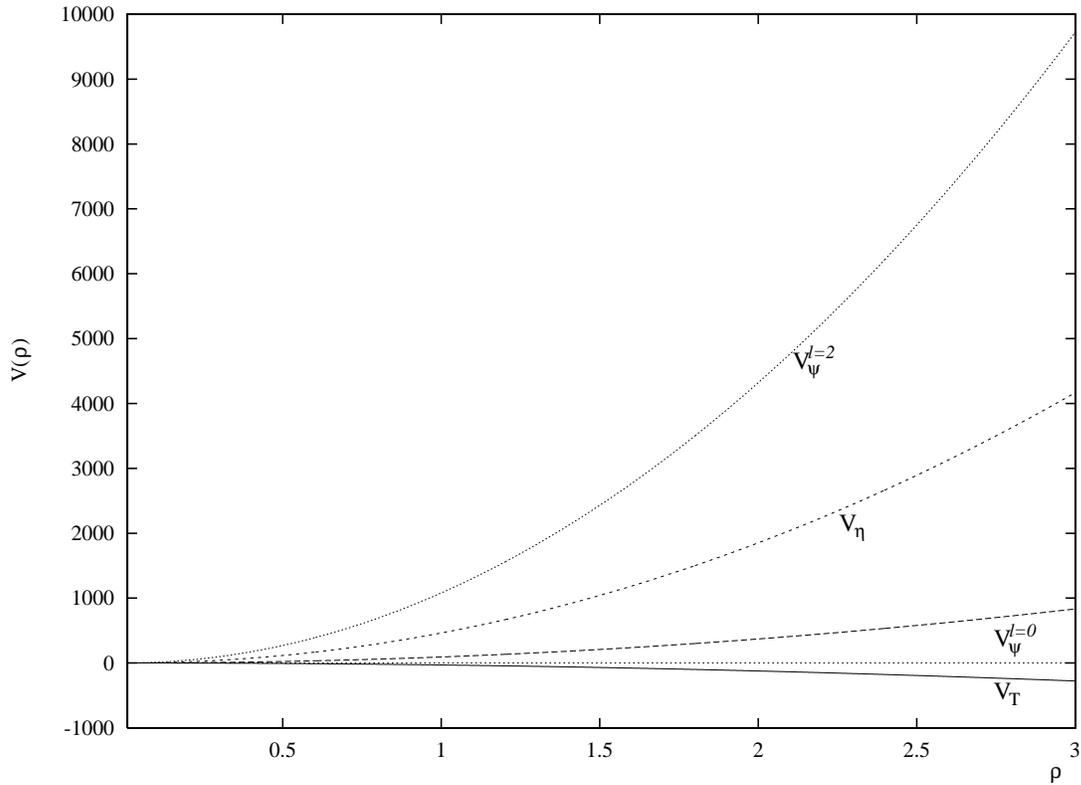,width=0.9\textwidth,clip=}
\caption{
\label{fig-potential-psi}
Four potential graphs ($V_{\rm T}, V_\psi^{l=0}, V_\eta, V_\psi^{l=2}$)
as functions of $\rho$ for an exact AdS$_3$.}
\end{figure}

Here we consider the geometry of an exact 
AdS$_3$(AdS$_3 \times $S$^3 \times $ T$^4$)
as the other solution to the type IIB string theory\cite{Lee98PRD104013}. 
This corresponds to the non-dilatonic solution.
This geometry can be led the BTZ black hole spacetime as a whole 
by the periodic identification.
A ten-dimensional minimally coupled scalar satisfies 
\begin{equation}
\Box_{10} \Psi =0.
\label{eq-minimal}
\end{equation}
$\Psi$ can be decomposed into
\begin{equation}
\Psi = e^{-i \omega t} e^{i m \varphi} e^{i K_i x^i}
          Y_l(\theta_1, \theta_2, \theta_3) \psi(\rho).
\label{psi-decomp}
\end{equation}
Then Eq.(\ref{eq-minimal}) leads to
\begin{equation}
\nabla^2_{\rm BTZ} \psi(\rho) + { \mu \over R^2} \psi(\rho) =0
\label{eq-BTZ}
\end{equation}
with $\mu= -l (l+2) - K^2 r_5^2$. 
The $\mu=-8(l=2)$ case contains both the dilaton($\Phi$) 
and a free scalar($\psi$) with $l=2$.
The $\mu=-3(l=1)$ case corresponds to an intermediate 
scalar ($\eta$) and $\mu=1$ leads to tachyon($T$).
Here the BTZ black hole spacetime is given by\cite{Ban93PRD1506}
\begin{equation}
ds^2_{\rm BTZ} = -f^2 dt^2 + 
    \rho^2 \left ( d \varphi - { J \over 2 \rho^2} dt \right ) ^2 +
    f^{-2} d \rho^2
\label{metric-BTZ}
\end{equation}
with $ f^2 = \rho^2/R^2 - M + J^2/4 \rho^2$
$=(\rho^2-\rho_+^2)(\rho^2-\rho_-^2)/\rho^2 R^2$. 
The mass, angular momentum, angular velocity at horizon 
and area of horizon are
\begin{equation}
M=(\rho_+^2 + \rho_-^2)/R^2, 
~~J = 2 \rho_+ \rho_- /R,
~~\Omega_H = { J \over 2 \rho_+^2},
~~{\cal A}_H^{\rm BTZ}=2 \pi \rho_+. 
\label{mass}
\end{equation}
Further one finds the relation between the BTZ and a 5D 
black holes as
\begin{equation}
T_H^{\rm BTZ} = (\rho_+^2- \rho_-^2)/2 \pi R^2 \rho_- = T_H, 
~~{1 \over T_{L/R}^{\rm BTZ}} = {1 \over T_H^{\rm BTZ}} 
 \left ( 1 \pm { \rho_+ \over \rho_-} \right ) = {1 \over T_{L/R}}.
\label{quantity-BTZ}
\end{equation}
Its s-wave equation with $m=0$ takes the form
\begin{eqnarray}
\left [ f^2 \partial_\rho^2
+ \left\{ {1 \over \rho} \partial_\rho \left ( \rho f^2\right ) \right\} 
     \partial_\rho
+{\omega^2 \over f^2} +{\mu \over R^2}
\right ] \psi(\rho) =0.
\label{eq-decoupled}
\end{eqnarray}
Defining $\psi(\rho) = \tilde \psi / \sqrt{\rho}$ and then 
(\ref{eq-decoupled}) takes the form 
\begin{equation}
f^2 \tilde \psi'' + (f^2)' \tilde \psi' +
\left [ { f^2 \over 4 \rho^2} - { (f^2)' \over 2 \rho} 
     + { \omega^2 \over f^2} + { \mu \over R^2} \right ]
  \tilde \psi =0,
\label{eq-dec-tilde}
\end{equation}
where the prime($'$) denotes the differentiation with respect to $\rho$.
In order to obtain the Schr\"odinger-type equation, we introduce 
the tortoise coordinate $\rho^*$ as\cite{Cha97PRD7546}
\begin{equation}
\rho^* = \int { d \rho \over f^2 } 
  = { R^2 \over 2 (\rho_+^2 - \rho_-^2)} 
   \left [ \rho_+ \ln \left ( {{\rho -\rho_+} \over {\rho +\rho_-}} \right ) 
   - \rho_- \ln  \left ( {{\rho -\rho_-} \over {\rho+ \rho_-}} \right )
   \right ].
\label{tortoise}
\end{equation}
We note that $\rho_+ \le \rho \le \infty$, while 
$-\infty \le \rho^* \le 0$ for outside horizon. 
On the other hand, in a 5D black hole one finds that $r_0 \le r \le \infty$ 
is mapped into $-\infty \le r^* \le \infty$.
Then (\ref{eq-dec-tilde}) leads to
\begin{equation}
 { d^2 \tilde \psi \over d {\rho^*}^2 } + 
 \left ( \omega^2 - V_\mu \right ) \tilde \psi =0,
\label{eq-sch}
\end{equation}
where the potential is given by
\begin{equation}
V_\mu(\rho) =  f^2 \left [ 
 - {f^2 \over 4 \rho^2} + {(f^2)' \over 2 \rho} 
 - { \mu \over R^2} \right ].
\label{potential-tortoise}
\end{equation}
Four potential graphs($V_{\rm T}, V_\psi^{l=0}, V_\eta, V_\psi^{l=2}$) 
in an exact AdS$_3$ background are shown in Fig.\ref{fig-potential-psi}. 
The parameters are chosen as $\rho_+=0.01, \rho_-=0.001, R=0.3$.
These are all monotonically increasing functions with $\rho$, 
contrast to $V_N$ for an AdS$_3$ bubble. 
This shows a peculiar property of AdS$_3$, which in this spacetime
the asymptotic states cannot be defined.
But the tachyon potential($V_T$) is a monotonically decreasing function. 
This field couples to the minimal weight primary operator 
with (1/2,1/2)\cite{Mal98JHEP12005}.
Also this satisfies both the stability condition for AdS$_3$ and the Dirichlet 
boundary condition\cite{Lee98PLB83}, which are clearly related to a shape 
of its potential.

\subsection{Asymptotically AdS$_3$-behavior and non-normalizable modes}
In the near-horizon($\rho \to \rho_+, V_\mu \to 0$), (\ref{eq-sch}) 
reduces to
\begin{equation}
{d^2 \tilde \psi_{\rm NH} \over d {\rho^*}^2 } +
   \omega^2 \tilde \psi_{\rm NH} =0
\label{eq-NH}
\end{equation}
which leads to the plane-wave solution
\begin{equation}
\tilde \psi_{\rm EH} = e^{-i \omega\rho^* } 
      + R_\psi^{\rm in}(\omega) e^{i \omega \rho^* } .
\label{plane-sol}
\end{equation}
Here the first term is an outgoing mode($\to$) and the second is 
an incoming mode($\leftarrow$).
Now let us discuss the asymptotically AdS$_3$-behavior.
Near the timelike boundary($\rho \to \infty, \rho^* \to 0$), 
one finds
\begin{equation}
{ d^2 \tilde \psi_\infty \over d {\rho^*}^2 } 
 - \left [ {3 \over 4} - \mu \right ] 
  { \rho^2 \over R^4} \tilde \psi_\infty =0 .
\label{eq-boundary}
\end{equation}
Here we introduce the relation between $\rho$ and $\rho^*$
\begin{equation}
{\rho_+ \over \rho(\rho^*)} = { Y \over 1 -\sigma^2} 
 \sum_{n=0}^\infty a_n(\sigma) Y^{2n}, 
 ~~Y = \tanh \lambda \rho^*, 
 ~~\sigma = { \rho_- \over \rho_+},
 ~~\lambda = { \rho_+ \over R^2} ( 1- \sigma^2),
 ~~a_0=1 .
\label{rho-relation}
\end{equation}
If $\sigma^2$ is very small and $\rho^* \to 0$, one finds
\begin{equation}
\rho = { \rho_+ \over Y} ( 1 - \sigma^2 ) =
     \rho_+ (1 - \sigma^2) \coth \lambda \rho^* \simeq 
    { R^2 \over \rho^*}.
\label{eq-rho}
\end{equation}
Using (\ref{eq-rho}), (\ref{eq-boundary}) leads to
\begin{equation}
{ d^2 \tilde \psi_\infty \over d{\rho^*}^2 } 
 - \left [ { 3 \over 4} - \mu \right ] {\tilde \psi_\infty \over {\rho^*}^2 } 
   = 0.
\label{eq-bnd-tortoise}
\end{equation}
Its solution takes the form
\begin{equation}
\tilde \psi_\infty = {\rho^*}^{{1 \pm 2 \sqrt{1 -\mu} } \over 2} .
\label{sol-boundary}
\end{equation}
Finally we have
\begin{equation}
\psi_\infty ( \rho^* ) = { \tilde \psi_\infty \over \sqrt{\rho} } 
 \propto {\rho^*}^{(1 \pm \sqrt{1-\mu})}.
\label{sol-bnd-final}
\end{equation}
For the s-wave free scalar($\mu=0$), $ \psi_\infty$ takes 
the form(${\rho^*}^2$, constant) and 
for the dilaton field($\mu=-8$), one finds 
(${\rho^*}^4, 1/{\rho^*}^2$). 
We find (${\rho^*}^{3/2}, {\rho^*}^{1/2}$) for a conformally 
coupled scalar($\mu=3/4$)\cite{Cha97PRD7546}.
For the tachyon($\mu=1$), one has ($\rho^*, \rho^*$) and for 
an intermediate scalar($\mu=-3$), one finds (${\rho^*}^3, 1/\rho^* $).
Instead of the plane-wave from, here one finds the 
power-law behaviors of $\rho^*$\cite{Bal99PRD046003}. The non-normalizable 
modes are found as $1/{\rho^*}^2$ for the 
dilaton and $1/\rho^*$ for an intermediate scalar, which 
diverge as $\rho^*$ approaches the timelike bounadry($\rho^*\to 0$).
The positive powers of $\rho^*$ all belong to be the 
normalizable modes, which converge at spatial infinity.
We note that the tachyon takes only the normalizable modes\cite{Lee98PLB83}.
This can be easily conjectured from its shape of potential $V_T$.
The s-wave free scalar takes a constant-behavior at infinity. 
This makes an difficulty in dividing $ \psi_\infty$ into 
an ingoing and outgoing modes at spatial infinity\cite{Bir97PLB281}.
Actually it is impossible to define an ingoing wave and an outgoing 
wave at spatial infinity of an exact AdS$_3$.
Instead in the asymptotically AdS$_3$ space 
it contains the normalizable as well as 
the non-normalizable modes. 
The latter will play an important role in calculation of 
the absorption coefficient ${\cal A}_\psi^{\rm in}$.

\subsection{Scattering from an exact AdS$_3$}
\begin{figure}
\epsfig{file=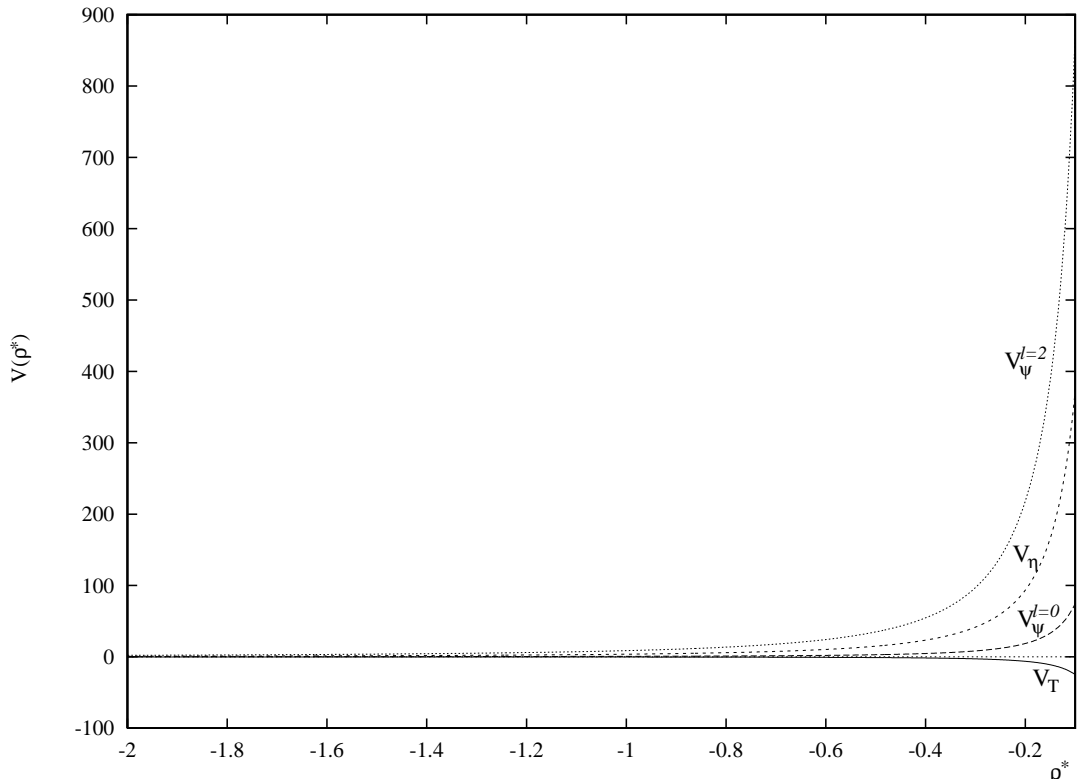,width=0.9\textwidth,clip=}
\vspace*{-12pt}
\caption{
\label{fig-potential-tortoise}
Four potential graphs ($V_{\rm T}, V_\psi^{l=0}, V_\eta, V_\psi^{l=2}$)
as functions of $\rho^*$ for an exact AdS$_3$.}
\end{figure}

First we note that $\rho^*$ covers only the 
left-hand side($-\infty \le \rho^* \le 0$) of the whole space. 
In the case of where the black hole geometry is asymptotically 
flat as a 5D black hole, the tortoise coordinate $r^*$ goes from 
$-\infty$ to $\infty$. Hence this is similar to the 
infinite string problem in which the initial data propagates towards 
left and right indefinitely\cite{Cha97PRD7546}. 
The initial data no longer enjoys this privilege when the background 
is asymptotically AdS$_3$ because the tortoise coordinate $\rho^*$ 
goes from $-\infty$ to 0 only. 
One may consider this as the semi-infinite string 
problem(or a finite cavity with reflecting 
walls in AdS$_5$\cite{Sus9901079}).
Then the Dirichlet or Neumann boundary condition at spatial 
infinity($\rho^*=0$) is required to formulate the problem appropriately.
However, we take a different point of view to attack 
an asymptotically AdS$_3$ problem.
This is based on the observation of the shape of potential and the global 
structure of an exact AdS$_3$\cite{Ban93PRD1506}.
We first construct the potential $V_\mu(\rho^*)$
by replacing $\rho$ in (\ref{potential-tortoise}) 
with $\rho \simeq \rho_+ \coth(\rho_+ \rho^*/R^2 )$.
The potentials in Fig.\ref{fig-potential-tortoise} look like 
the half of the AdS$_3$-bubble potentials.
Especially, we observe that the free field potentials $V_\psi^{l=0}(\rho^*)$ 
and $V_\psi^{l=2}(\rho^*)$ in Fig.\ref{fig-potential-tortoise} take 
nearly the same form as in $V_\phi^{l=0}(r^*)$ and 
($V_\phi^{l=2}(r^*), V_\nu(r^*), V_\lambda(r^*)$) in the region of 
$-\infty \le r^* \le 0$ in Fig.\ref{fig-potential-AFS-tor}.
Furthermore the Penrose diagram of an exact AdS$_3$($-\infty \le \rho^* \le 0$) 
is an half of the would-be whole diagram in 
$-\infty \le \rho^* \le \infty$\cite{Ban93PRD1506}.

In this work we assume that the potential of an exact AdS$_3$ 
is the left-hand side one of an AdS$_3$ bubble.
The important thing is to calculate the absorption 
cross section in the background of an exact AdS$_3$.
Considering the $\{ {\rm in}\}_\psi$-state picture, it is not hard to 
calculate the absorption coefficient ${\cal A}_\psi^{\rm in}$.
Although the $\{ {\rm out} \}_\psi$-state cannot be defined, 
we can derive ${\cal A}_\psi^{\rm in}$ from 
the backscattering of a test field $\psi$ off 
$V_\psi(\rho^*)$.
If we choose the boundary condition appropriately, the 
of potential Fig.\ref{fig-potential-tortoise} is 
enough to calculate the absorption coefficient, 
regardless of the would-be right-hand side($ 0 \le \rho^* \le \infty$).
In this backscattering process, 
the key point is to use an appropriate matching procedure 
between a near AdS$_3$ and an asymptotically AdS$_3$.
We remind the reader that 
$\rho^*=0$ is a timelike bounadry and thus information enter or exit 
from it. This is an exact middle point if one assumes a whole 
space of $-\infty \le \rho^* \le \infty$. Thus requiring the conventional 
boundary condition may lead to a wrong result in deriving 
the absorption coefficient.
Instead of the Dirichlet condition of $\psi \vert_{\rho^*=0} =0$, 
one may use the non-normalizable modes. 
The non-normalizable mode is a divergent quantity at $\rho^*=0$ 
but its flux is finite at $\rho^*=0$.
Also it corresponds to specifying another 
boundary condition at spatial infinity.
On the other hand, 
if one use the normalizable modes which satisfy the Dirichlet 
boundary condition, one may not succeed in obtaining 
the absorption coefficient in an exact AdS$_3$.

\subsection{AdS$_3$-AdS$_3$ matching procedure and absorption cross section}
In order to calculate the semiclassical absorption cross section, we have 
to solve the exact differential equation Eq.(\ref{eq-BTZ}) with 
an appropriate boundary condition. Since it is difficult
to solve Eq.(\ref{eq-BTZ}) directly, one has to use the 
matching prodecure between the near-horizon($\rho \sim \rho_+$) AdS$_3$ and 
the far-region($\rho \to \infty$) AdS$_3$.
Here the matching point resides on $ -\infty < \rho^* < 0$ 
in Fig.\ref{fig-potential-tortoise}.

In the far region Eq.(\ref{eq-BTZ}) becomes
\begin{equation}
\psi''_\infty + { 3 \over x} \psi'_\infty +
  { \mu \over x^2} \psi_\infty =0
\label{eq-far}
\end{equation}
with a dimensionless variable $x = \rho/R$.
One easily finds the far-region solution
\begin{equation}
\psi_\infty ( x) = \left [ \tilde \alpha x^{-1 + \sqrt{1-\mu}} 
  + \tilde \beta x^{-1 - \sqrt{1-\mu} } \right ]
\label{sol-far}
\end{equation}
with two unknown constants $\tilde \alpha, \tilde \beta$.
The first term is a divergent quantity at $\rho=\infty$ but 
behaves well in the interior region.
This corresponds to the non-normalizable modes and is 
coupled to the boundary operator ${\cal O}$ at infinity. 
The second one is the normalizable mode and propagates 
in the bulk. This can be used to construct two, three and four-point 
functions.
Although one cannot define an ingoing flux at infinity 
of AdS$_3$, one can calculate the total flux.
The flux at spatial infinity is given by 
\begin{equation}
{\cal F}(\infty) = -2 \pi \sqrt{1-\mu}
\left \vert \tilde \alpha - i \tilde \beta \right \vert^2.
\label{influx-far}
\end{equation}
We note here that for the tachyon with $\mu=1$, 
${\cal F}^T(\infty)=0$. This is because the tachyon 
takes only the normalizable modes which are to be zero at 
infinity. Thus we exclude it from our analysis.

In order to obatin the near-horizon behavior, 
we introduce the new variable 
$z=(\rho^2 -\rho_+^2)/(\rho^2 - \rho_-^2)$
$=(x^2 - x_+^2) /( x^2 - x_-^2)$.
Then Eq.(\ref{eq-BTZ}) leads to
\begin{equation}
z(1-z) {{d^2 \psi} \over dz^2}
+(1-z) {{d \psi} \over d z}
+\left ( {A_1 \over z} +{{\mu/4} \over 1-z} -B_1 \right ) \psi
=0,
\label{eq-hyper}
\end{equation}
where
\begin{equation}
A_1 =
\left ( {{\omega - m \Omega_H} \over 4 \pi T_H} \right )^2,
~~B_1 = - {\rho_-^2 \over \rho_+^2}
\left ({{\omega - m \Omega_H \rho_+^2 / \rho_-^2} \over
4 \pi T_H^{\rm BTZ}} \right )^2.
\label{A1B1}
\end{equation}
In the case of the s-wave propagation with $m=0$, the 
near-horizon AdS$_3$ equation (\ref{eq-hyper}) leads exactly to 
the near-horizon equation (\ref{near-eq}) of a 5D black hole.
Explicitly, the relationship between these is given by
\begin{equation}
A_1 \to Q, ~~{\mu \over 4} \to E, ~~B_1 \to C.
\label{relation}
\end{equation}
In this sense, (\ref{near-eq}) is called an AdS$_3$ bubble.
The ingoing wave is given by the hypergeometric function
\begin{eqnarray}
\psi(z) &=&
C_1 z^{-i \sqrt{A_1}} (1 -z )^{(1 - \sqrt{1-\mu})} F(a,b,c;z),
\label{sol-hyper}
\end{eqnarray}
where
\begin{eqnarray}
&&\hspace*{-2em}\sqrt{1-\mu} \simeq l+1, 
\label{mu-l} \\
a&=& {1 - \sqrt{1-\mu} \over 2} - i \sqrt{A_1} +i\sqrt{B_1} 
\simeq - { l \over 2} - i { \omega \over 4 \pi T_R^{\rm BTZ}},
\label{hyper-a1} \\
b&=& {1 - \sqrt{1-\mu} \over 2} - i \sqrt{A_1} -i\sqrt{B_1} 
\simeq - { l \over 2} - i { \omega \over 4 \pi T_L^{\rm BTZ}},
\label{hyper-b1} \\
c&=& 1 - 2 i \sqrt{A_1}= 1 - i { \omega \over 2 \pi T_H^{\rm BTZ}} .
\label{hyper-c1}
\end{eqnarray}
Here $\simeq$ means $K^2 r_5^2 \simeq 0$.
The coressponding flux is
\begin{equation}
{\cal F}(0) = -8 \pi \sqrt{A_1} (x_+^2 - x_-^2) 
\left \vert C_1 \right \vert^2
\label{influx-near}
\end{equation}
with $x_+^2 - x_-^2 = (r_0/R)^2 \ll 1$.
The absorption coefficient will be taken as
\begin{equation}
{\cal A}_\psi^{\rm in} =
{{\cal F}(0) \over {\cal F}(\infty)} =
{ 4 \sqrt{A_1} (x_+^2-x_-^2) \over \sqrt{1-\mu} }
{|C_1|^2 \over |\tilde \alpha - i \tilde \beta |^2}.
\label{abs-psi}
\end{equation}
In order to obtain $\tilde \alpha$ and $\tilde \beta$, we 
use the matching procedure. It is important to remember that the 
present spacetime is an exact AdS$_3$. Thus we have to match 
the near-AdS$_3$ with asymptotically AdS$_3$ to find an absorption 
coefficient.
We know the far-region behavior of (\ref{sol-hyper}). 
This can be found from the $z \to 1-z$ for the 
hypegeometric function
\begin{eqnarray}
\psi_{n \to f} (x) &=& 
\left [ 
C_1 E_1 ( x_+^2 - x_-^2 )^{(1-\sqrt{1-\mu})/2} x^{-1 + \sqrt{1-\mu} }
+ C_1 E_2 ( x_+^2 - x_-^2 )^{(1+\sqrt{1-\mu})/2} x^{-1 - \sqrt{1-\mu} }
\right ],
\label{sol-hyp-far}
\end{eqnarray}
where 
\begin{equation}
E_1 = { \Gamma(c) \Gamma(c-a-b) \over \Gamma(c-a) \Gamma(c-b) },
E_2 = { \Gamma(c) \Gamma(-c+a+b) \over \Gamma(a) \Gamma(b) }.
\label{E1E2}
\end{equation}
Matching (\ref{sol-far}) with (\ref{sol-hyp-far}) in the 
far-region($ x \gg 1$) leads to
\begin{equation}
\tilde \alpha = C_1 E_1 \left ( { r_0 \over R} \right ) ^{1 -\sqrt{1-\mu}},
\tilde \beta = C_1 E_2 \left ( { r_0 \over R} \right ) ^{1 +\sqrt{1-\mu}}.
\label{coef-alphabeta}
\end{equation} 
Considering $R \gg r_0$, one finds $\tilde \alpha \gg \tilde \beta$ even 
for $l=0$ case. Hence we can neglect $\tilde \beta$ in favor of $\tilde \alpha$.
This amounts to taking the flux of the non-normalizable modes.
The key point is an AdS$_3$-AdS$_3$ matching in this backscattering process.
Then the absorption coefficient is approximately given by
\begin{equation}
{\cal A}_\psi \simeq 
{ 4 \sqrt{A_1} (x_+^2 - x_-^2) \over \sqrt{1 -\mu} } 
\left ( { r_0 \over R} \right ) ^{2 (\sqrt{1-\mu} -1 )}
{ 1 \over |E_1|^2 }.
\label{abs-psi-match}
\end{equation}
The absorption cross section for AdS$_3\times$S$^3$ with $m=0$ leads to
\begin{eqnarray}
 \sigma^{\psi}_{\rm AdS} 
&=&  
{{\cal A}_\psi \over \omega}
\simeq 
{\tilde {\cal A}_H^{\rm 6D} \over l! (l+1)! } 
\left ( { r_0 \over R } \right )^{2l} 
\left \vert
{
{\Gamma({l+2 \over 2} - i { \omega \over 4 \pi T_L^{\rm BTZ} } ) 
 \Gamma({l+2 \over 2} - i { \omega \over 4 \pi T_R^{\rm BTZ} } )} 
\over
 \Gamma(l - i { \omega \over 2 \pi T_H^{\rm BTZ} } )} 
\right \vert ^2
\label{abs-ads}
\end{eqnarray}
with $\tilde {\cal A}_H^{\rm 6D} = {\cal A}_H^{\rm BTZ} \times 2 \pi^2 R^3$.

In the low energy limit $\omega \to 0$,
it turns out that the 6D cross sections for
an exact AdS-theory take the same form
as (\ref{sigma-phi0}) and (\ref{sigma-nu})
\begin{eqnarray}
&& \sigma_{\rm AdS}^{\psi_0}  = \tilde {\cal A}_H^{\rm 6D},
\label{sigma-psi0} \\
&&\sigma_{\rm AdS}^{\psi_2}  = \sigma_{\rm AdS}^{\Phi}  =
{1 \over 3}{\tilde {\cal A}_H^{\rm 6D} \over 4}\left ({r_0 \over R}\right )^4 .
\label{sigma-psi2}
\end{eqnarray}
This 6D result is derived from AdS$_3\times$S$^3$.
In order for this to match with the cross section of a 5D black hole, 
it needs to introduce a compactified circle(S$^1$) in 
M$_5\times$S$^1\times$T$^4$.
In this case one finds $\tilde {\cal A}_H^{\rm 6D} = {\cal A}_H^{\rm 5D} 
\times 2 \pi R$ with a radius of S$^1$($R$).
We note that $\phi_2$, $\nu(=\Phi)$ and $\lambda$ give us  
slightly different cross sections in an AdS$_3$ bubble,
whereas these($\psi_2, \Phi$) 
do not make any distinction in an exact AdS-theory.

\section{Discussions}
It seems that the S-matrix cannot be extracted from the anti-de Sitter 
space even in a limit\cite{Sus9901079,Pol9901076}. 
This is based on the fact that in an 
exact AdS$_3$ the asymptotic states cannot be defined, due to 
the timelike bounadry and the periodicity of geodesics.
However, the authors in \cite{Bal99JHEP03001} showed 
that the correlation functions of the 
dual CFT$_4$ to AdS$_5$ are considered as the bulk 
S-matrices. The vacuum correlators 
$\langle {\cal O}(x_1){\cal O}(x_2)\cdots {\cal O}(x_n)\rangle_{{\rm CFT}_4}$
of the CFT$_4$ are expressed 
as truncated n-point functions convolved against 
the non-normalizable modes.
These can be interpreted as an S-matrix for an exact AdS$_5$ arising 
from a limit of scattering from an AdS$_5$ bubble in asymptotically 
flat space.

In this work, we show that the S-matrix of an AdS$_3$ bubble 
can be derived from an exact AdS$_3$ in the dilute gas and low 
energy limits.
We confirm this from the calculation of the absorption cross section. 
This originates from the fact that the near-horizon equations
for an AdS$_3$ bubble (\ref{near-eq}) and an exact
AdS$_3$ (\ref{eq-hyper}) are the same 
form, but they have different boundary conditions at infinity.
In the AdS$_3$ bubble-calculation, 
one uses the AdS$_3$-AFS matching to obtain the absorption coefficient 
${\cal A}_N^{\rm out}$. On the other hand, in the exact AdS$_3$-calculation 
we use both the \{in\}-state 
picture and the non-normalizable mode to obtain 
the absorption coefficient ${\cal A}_\psi^{\rm in}$.
This amounts to taking the AdS$_3$-AdS$_3$ matching.
The s-wave greybody factor of a free scalar of an AdS$_3$ bubble 
has exactly the same form as that of an exact AdS$_3$. 
For the dilaton we find the same form of cross section 
$\sigma = c \tilde {\cal A}_H^{\rm 6D} (r_0/R)^4$ 
but with $c=1/4$ for an AdS$_3$ bubble and 
$c=1/12$ for an exact AdS$_3$.

Let us compare our results with the others.
The general formular for the gerybody factor is derived from 
the vacuum two-point function 
$\langle {\cal O}(x){\cal O}(0) \rangle_{{\rm CFT}_2}$
of a boundary operator ${\cal O}$ in the effective string\cite{Gub97PRD7854} 
and boundary CFT$_2$ approaches\cite{Teo98PLB269}.
These give us the same result for a free scalar but for the dilaton, 
$c=1/4$ as in an AdS$_3$ bubble.
Consequently, the two-point correlator provides us the 
greybody factor in the dilute gas and low energy limits.
This quantity takes exactly the same form in the CFT$_2$ and 
AdS$_3$ bubble approaches.
Further, in the exact AdS$_3$ approach one finds the same form 
of the greybody factor.
This means that the S-matrix can be derived from an exact AdS$_3$ space.
It is obvious that the conformal limit of the gauge theory(CFT$_2$) corresponds 
with scattering from an exact AdS$_3$.
Finally we present here a scattering picture in an exact AdS$_3$ and 
compare it with the scattering of an AdS$_3$ bubble in AFS.

\acknowledgments
This work was supported in part by the Basic Science Research Institute 
Program, Ministry of Education, Project NOs. BSRI-98-2413 and BSRI-98-2441.


\end{document}